\documentclass[11pt]{article}
\begin{document}
\author{ {\bf Jerzy Szwed}\footnote {Work supported in part by the Marie Curie Actions Transfer of
Knowledge project COCOS (contract MTKD-CT-2004-517186).}
\\
Institute of Physics, Jagellonian University,
\\ 
Reymonta 4, 30-059 Krak\'ow, Poland
}
\title{\Large \bf  Possible new wave phenomena in the brain.}
\maketitle

\begin{abstract}

We propose to search for new wave phenomena in the brain by using interference effects 
in analogy to the well-known double slit (Young) experiment. This method is able to extend the 
range of oscillation frequencies to much higher values than currently accessible. 
It is argued that such experiments may test the hypothesis of the wave nature of 
information coding.
\end{abstract}

1. 
Oscillatory effects in the brain have been known for more than a century and
intensively studied since then \cite{review book}. Although classified in detail 
and related to other brain functions, 
their exact role is still not well understood. 
In this context we point to the current discussion on possible new effects playing an  
important role in neural processes in the brain. 
Among the most spectacular hypotheses, the suggestion of non-algorithmic processes taking place in the mind, 
or the quantum nature of information-processing, are maybe
 the most spectacular ones \cite{Penrose}. 
In this note we propose a series 
of experiments testing the possibility of high-frequency wave phenomena not detectable by standard 
measurement methods. If found, the short wavelengths have the potential to play a crucial role in 
information-processing.
The proposed experiments are motivated by
their physics analogues, where they have contributed crucially to the development of our understanding 
of physics in the last few centuries.

2. When we look at the physical parameters of the known brain waves, 
it is hard to conclude what their role is in mental processes. 
With the measured oscillation frequencies in a rough range of  1 to 100 Hz (delta to gamma) and propagation 
velocities ranging from 0.5 
to 150 m/s, we arrive 
at wavelengths within the limits of 5 millimetres to 150 metres. This means that the waves measured 
at present, while reflecting some collective phenomena, are too long to be directly responsible for  
  information-coding in the brain. The recently studied ultra-high frequency
 oscillations --- up to $1000$ Hz ---
cannot touch the interesting short wave length region either, even   if they propagated with very low
 velocities. The proposed experimental scheme makes it possible to study a new range of frequencies
 and corresponding wavelengths. It is sometimes believed that neural processes take place on a millisecond scale.
 This would mean that a search for oscillations of frequencies higher than 1000 Hz cannot bring
 positive results. But a firm answer to this question can be given only by experiment.

3. The fundamental experiment proving the wave nature of light was performed by Thomas Young \cite{Young}. 
The experimental setup, formulated in more general terms, 
consists of a single signal source  
  and two slits (gates) between the source and target. These gates in fact form two separate, coherent signal sources. Covering one of the two slits  
 the signal travels through the second slit 
and hits the target producing a picture opposite the opened slit. Analogically, covering the
 second slit we observe the picture on the target opposite the first slit. 
With both slits opened a spectacular result is observed. It is not a simple 
sum of both previous pictures, but instead an interference pattern of dark and light bands that
appears. Moreover, at maximum brightness, the signal is not just twice as strong (as one would expect from summing the two separate signals) but 4 times stronger. 
Today the explanation is well known --- what is travelling between the source and screen are waves. In the target these waves meet, 
having covered in general a different distance when travelling through the first and second slit. If the path difference is equal to the integer number of wavelengths --- 
the waves interfere positively and produce a maximum; in the other extreme case when the path difference is half the integer number of wavelengths --- 
the waves interfere negatively 
and produce a minimum. The result in the target is a characteristic interference pattern. 
The second important effect is the strength of the signal. 
In most double slit experiments we measure the energy deposited on the screen, and this energy is proportional to the square of the incoming wave
 (this is also so with the quantum case). 
Therefore, when two equal waves meet "in phase" the effective wave is twice as high but the signal strength is 4 times stronger. 
In some experiments one can measure the wave directly, e.g. the electric potential, then the abnormal amplification does not take place. 
Finally one should stress
an important detail of the experimental setup which makes the observation possible. The interference pattern of waves is extended in space due to the 
large distance $L$ between the slits and target as compared to the slit separation $d$ --- approximately by a factor of
$L/d$ . This is why we do not need the accuracy of the order of the wavelength $\lambda$, but rather $\lambda L/d$. Choosing
the slit separation $d$ and the distance 'slits--target' $L$ properly, we are able to see interference taking place at an inaccessibly small scale 
 magnified to measurable distances at the target.

The original double slit experiment was repeated more than a century later in its new version, contributing crucially to our understanding of quantum mechanics. 
In 1909 G.I. Taylor used single photons as signal source \cite{Taylor}. Analogical experiments with "quantum waves" 
of matter were performed with electrons \cite{Jonsson} in 1961, single electrons \cite{Tonomura} in 1989, neutrons \cite{Zeilinger} in 1988, helium atoms \cite{Mlynek} in 
1991 and  fullerene (C60) molecules \cite{Arndt} in 1999.

4. The aim of this note is to propose 
an analogical experiment in the brain. We do not attempt to join the speculations as to what could be the origin of wave phenomena and whethe they are classical or quantum-like.  
Instead we would like to concentrate on the experimental feasibility of such experiments. In general, we should investigate the spatial and/or temporal structure 
of effective 
signals resulting from two different sources. 
The main unknown which determines the experimental setup is the scale (in distance and in time) at which the phenomenon manifests itself. Therefore, one is 
forced to perform tests at levels ranging from intercellular (nm) distances, through single neuron ($\mu m$) to multi-neuron (mm) distances and corresponding time scales. 
These scales determine the signal source and the target. 

There are two general setups leading to possible interference effects.  
The first one, which is copied exactly from the Young experiment, we call "spatial" setup. A single source (or two correlated/coherent sources) emits 
the signal, which is carried to the target via two distinct ways. Before hitting the target both signal carriers mix in some overlapping volume, 
where they possibly interfere. The target is spatially large enough that our detectors distinguish details within it (the space resolution of 
our detector should be much smaller than the area reached by the signal). As in the Young experiment, due to the different distances covered by the two signal carriers,
the effective signal will show enhanced and suppressed interference bands as we move across the target. 
Moreover, the strength of the enhanced signal can be higher than the simple sum of two single signals in the
 case when we measure the "energy deposition" (classical) or events (quantum) rather than the wave itself.

The second setup uses a time difference rather than a path difference, and we therefore call it a "temporal" setup. It assumes two sources which can emit signals at 
different (in general), well-controlled instants. The detector in target, where the effective signal is measured, can be simpler than in the spatial setup – 
it can be just a single measuring unit. The possible 
interference effect will be observed by shifting the relative time of emission of the two signals. The characteristic enhancement-suppression pattern emerges here 
as a function of time difference. Compared to the "spatial" case the "temporal" setup looks simpler, but requires more precision. 
The resolution should be high enough that we are able to distinguish details at times comparable to the wave period. In the "spatial" case 
the interference pattern can be extended in space due to the magnification factor mentioned above.

Both presented setups can be refined or combined to make the search as complete as possible. This is especially important due to our lack of knowledge on what the scale 
(wavelength, frequency) of the phenomenon is. 
The possible quantum-like behaviour requires even more attention --- a systematic build-up of statistics.

5. How can we look for interference effects in the brain? We suggest a few examples on various scales, but the list is by no means complete.  
The signal sources can be relatively easily defined, they
 can be external (sensory, vision, sound, ...) 
or internal (direct stimulation of brain cells) – the control over their relative correlation is very important. 
 The correct preparation of target and removal, as much as possible, of side effects will be a real challenge. 

Perhaps the best place to perform the experiment is the brain slice. In the case of a "spatial" experiment it is easy to choose and adjust the distance $d$ between the
 two signal sources.
 and place the target --- a multi-electrode system in the form of spatial matrix --- at the other end of the slice at distance $L$. The wavelengths and frequencies 
accessible in this setup can cover a broad range. Assuming for example the ratio $d/L \sim 0.1$ one can obtain wavelengths reduced by the same ratio (of the order of $10^{-4}m$) and
 frequencies in the range $10^{-4}$ --- $10^{-6}$. Playing with the parameters of the experiment $d$ and $L$, as well as with the signal velocity, one can easily go beyond this range. 
The "temporal" set-up in the brain slice is equally easy to arrange.

Another very good set-up  
 ready for both "spatial" and "temporal" experiments is the sensory system of the rat. The signal sources here are the rat's vibrissa and the target --- 
the barrel cortex. To facilitate the detection of the interference pattern the signal sources should preferably be of equal strength. 
In order to perform the "spatial" experiment one again needs a multi-electrode detector which allows for spatial resolution of effective signals. 
The scale at which the effect is searched can be chosen by the magnitude and separation of electrodes, and ranges from single neurons to multi-neuron systems. 
The "temporal" experiment seems slightly simpler: it can be performed with a single electrode but the time (in particular time shift) control must be very accurate. 
Performing the experiment \it in vivo \rm  one is able to experiment with the brain at rest or in an active state. 
Similar setups can be arranged with the visionary system. The signal sources can be 2 light signals separated in space and/or time. Again the right choice of the target 
 and multi-electrode detector would be crucial.

Going down in scale one can test the signal transmission inside the neuron. Although this process seems to be understood at electrochemical level
 some novel hypotheses \cite{Penrose} 
may be verified in this system. Coherence of the signal emission and transmission could be the main problems here.
Increasing the scale one may use such techniques as EEG, MEG or high-precision MRI 
to experiment at multi-neuronal systems. These devices allow for a wide choice of sources and targets; their resolution is however much lower. 

6. To summarize, this note suggests a systematic search for wave effects in the brain. Following the spectacular double slit interference 
experiment, which was successful 
so many times and in so many systems, we point out the most characteristic features of the interference pattern. 
 Without knowing the scale of possible effects it is reasonable to test as many experimental systems as possible where
  multi-stimulus sources and spatio-temporal extended targets are available. 
Since we are dealing with a very complex system, the removal of side effects so that we are as close as possible to the original 
two-slit experiment will probably be the hardest task.
However, the proposed experiments are in our opinion very important; they lead us to a range of oscillations inaccessible with standard methods.
In the case of a positive result --- the discovery of very short brain waves may shed new light on information-processing in the brain.

7. The author would like to thank Marian Lewandowski for discussions.


\end{document}